\documentstyle[aps,prbbib,twocolumn,epsf]{revtex}

\begin{document}

\draft
\title{Spin-current induced electric field} 

\author{Qing-feng Sun$^{1,2}$, Hong Guo$^{1,2}$, and Jian Wang$^3$}

\address{$^1$Center for the Physics of Materials and Department
of Physics, McGill University, Montreal, PQ, Canada H3A 2T8 \\
$^2$International Center for Quantum Structures, Institute of Physics,
Chinese Academy of Sciences, Beijing, China \\
$^3$Department of Physics, The University of Hong Kong, Pokfulam Rood,
Hong Kong, China
}

\maketitle

\begin{abstract}
We theoretically predict that a pure steady state spin-current without 
charge-current can induce an electric field. A formula for the induced
electric field is derived and we investigate its characteristics. 
Conversely, a moving spin is affected by an external electric field and 
we present a formula for the interaction energy.
\end{abstract}
\pacs{72.25.-b, 03.50.De}

In a traditional electric circuit the number of spin-up and spin-down 
electrons are the same, and both kinds of electrons move in the same direction
under an external electric field. The total spin-current
$I_s=\sigma(I_{\uparrow}-I_{\downarrow})$ is therefore zero, and only the 
charge-current $I_{ec}=e(I_{\uparrow}+I_{\downarrow})$ is relevant. When a system 
includes ferromagnetic materials or under an external magnetic field, electron 
spins can be polarized so that the total spin of the system is non-zero. 
Then, the corresponding charge-current is polarized, {\it i.e.} current due to 
spin-up electrons, $I_{\uparrow}$, is not equal to the spin-down current 
$I_{\downarrow}$, although both kinds of electrons move in the same direction, 
as schematically shown in Fig.1b. This gives a non-zero total spin-current. 
Spin-polarized charge-current has been the subject of extensive investigations
for last two decades.\cite{ref1,ref2} Recently, a very interesting extreme 
case of a finite spin-current without charge-current has been investigated by 
several groups\cite{ref3,ref4,ref5}. Such a situation comes about when spin-up 
electrons move to one direction while an equal number of spin-down electrons 
move to the opposite direction, as schematically shown in Fig.1a. Then the 
total charge-current is identically zero and only a net spin-current exists. 
This is just the opposite situation of the traditional charge-current without 
any spin. 

By Ampere's law, a charge-current induces a magnetic field in the space around 
it. In this paper, we ask and answer the following question: can a pure steady 
state spin-current without charge-current induce an electric field? 
The problem can be viewed in the following way. Associated with the electron
spin ${\bf \sigma}$, there is a magnetic moment $g\mu_B {\bf \sigma}$, where
$\mu_B$ is the Bohr magneton and $g$ is a constant. Therefore when there is
a spin-current $I_s$, there is a corresponding magnetic moment current 
$I_m =g\mu_B I_s$. In the rest of the paper, we theoretically prove that a 
magnetic moment current can induce an electric field. We further prove that
an external electric field can also act on a spin-current.

To start, we recall that a static classical magnetic moment ${\bf m}$ produces 
a magnetic field ${\bf B}$. Consider a classical magnetic moment ${\bf m}$ 
due to a tiny charge-current ring, see Fig.1c. The charge-current is $I_{ec}$ 
and radius of the ring is $\delta$. The magnetic field ${\bf B}$ of this 
charge-current ring is easily obtained by the Biot-Savart law. Then, let 
$\delta\rightarrow 0^+$ and $I_{ec}\rightarrow \infty$ but keep
${\bf m} =\pi \delta^2 I_{ec} \hat{n}_m$ as a constant ($\hat{n}_m$ is the 
unit vector of the magnetic moment), the magnetic field ${\bf B}$ 
due to magnetic moment ${\bf m}$, at space point ${\bf R}$, can be written as:
\begin{equation}
{\bf B} = -\nabla \frac{ \mu_0 {\bf m} \bullet {\bf R} } {4\pi R^3}\ \  .
\label{B0}
\end{equation}
Another method for obtaining the same magnetic field is by using the 
mathematical construction of equivalent magnetic ``charge''\cite{book}. In 
this method, we imagine the magnetic moment ${\bf m}$ being consisted of a 
positive and a negative magnetic ``charge'' $\pm q_{mc}$ situated very close 
to each other with a distance $\delta$ (see Fig.1d). When 
$\delta\rightarrow 0^+$ and $q_{mc}\rightarrow \infty$, we hold 
${\bf m}=q_{mc} \delta \hat{n}_m$ as a constant. Each magnetic ``charge'' 
$q_{mc}$ produces a magnetic field $\frac{\mu_0 q_{mc} {\bf R} }{4\pi R^3} $.
The field ${\bf B}$ induced by magnetic moment ${\bf m}$ can then
be obtained by adding contributions of the two magnetic ``charges'' 
$\pm q_{mc}$. Of course, we again obtain Eq.(\ref{B0}). Note that the 
language of magnetic ``charge'' is only a mathematical construction convenient
for our derivations\cite{book}, and no magnetic monopole is hinted whatsoever.

After reviewing the magnetic field of a {\it static} magnetic moment,
in the rest of the paper we consider magnetic moments in motion. In the first
example we consider the simplest case of a classical infinitely long 
one-dimensional (1D) lattice of chargeless magnetic moment, with the whole 
lattice moving with speed ${\bf v}$ (see Fig.1e). This gives a pure 
steady state magnetic moment current. Since nothing is changing with 
time, a very surprising result, as we now prove, is that this magnetic 
moment current induces an {\it electric} field.
The induced field for this situation can be calculated exactly by a simple 
Lorentz transform, therefore providing a benchmark result for our more general 
results to be discussed later. Let $\rho_m$ to be the linear density of magnetic 
moment for the lattice and we will use two methods to solve the electromagnetic 
field of the spin-current.\\
{\bf Method One.} We first solve the total {\it magnetic} field of a 
{\it static} 1D magnetic moment lattice by integrating Eq.(\ref{B0}) over 
the lattice. This is easy to do and we call the result ${\bf B}_{static}$:
$
{\bf B}_{static}
= -\nabla  \frac{ \mu_0 \rho_m \hat{n}_m  \bullet {\bf R}_{\perp} }
{2\pi R_{\perp}^2 },
$
where ${\bf R}_{\perp} ={\bf R}-({\bf R} \bullet \hat{l})$. Here $\hat{l}$ is 
unit vector along the lattice. Then we make a relativistic transformation: 
the electromagnetic field of the moving magnetic moment lattice can be 
obtained straightforwardly by the Lorentz transform of ${\bf B}_{static}$. 
The results are
\begin{eqnarray}
{\bf B} &=& \gamma {\bf B}_{static}\  , \\
\label{B1}
{\bf E} & = &-\gamma {\bf v}\times {\bf B}_{static}\ , 
\label{E1}
\end{eqnarray}
where $\gamma =1/\sqrt{1-\frac{v^2}{c^2}}$.
Clearly, 
we have an induced electric field ${\bf E}$ and this field 
is related to ${\bf v}$. We note that although the results are 
unambiguously obtained, we have not identified the physical origin 
of the resulting electromagnetic field, {\it i.e.} this method does not tell 
us whether the field is induced by the magnetic moment or by the magnetic 
moment current. For this reason, we analyze the same problem again from a 
second method.\\
{\bf Method Two.} 
Here we use the equivalent magnetic ``charge'' method discussed
above.
This means removing the current density of the ring at the equation 
$\nabla \times {\bf B} =\mu_0 {\bf J}_{ec} +\mu_0\epsilon_0 
\frac{\partial {\bf E}}{\partial t}$, and adding the imaginary magnetic 
charge at the equation $\nabla \bullet {\bf B} =0$, {\it i.e.} this equation
changes to $\nabla \bullet {\bf B} =\mu_0 \rho_{mc}$, where $\rho_{mc}$ is the
volume density of magnetic ``charge''. We emphasize again that this practice 
is only a mathematical trick to solve our problem. When our magnetic 
moment moves, the original Maxwell equation in which the magnetic moment 
${\bf m}$ is described by a tiny charge-current ring, satisfies
relativistic covariance. Clearly, Maxwell equations after the equivalent 
magnetic charge transformation must also satisfy relativistic covariance. 
This covariance can be achieved, as shown in standard textbook\cite{book}, 
by changing the Maxwell equation to
$\nabla \times {\bf E} = -\frac{\partial {\bf B}}{\partial t}$ to
$\nabla \times {\bf E} = -\frac{\partial {\bf B}}{\partial t} -\mu_0 
{\bf J}_{mc}$, where ${\bf J}_{mc}$ is the magnetic ``charge'' 
current.\cite{book} The last equation means that a moving magnetic ``charge'' 
can produce an electric field. The electric field ${\bf E}$ produced by a 
volume (linear) element of magnetic ``charge'' current, 
${\bf J}_{mc} dV$ ($I_{mc} d{\bf l}$), is simply:
\begin{equation}
{\bf E} =-\frac{\mu_0 {\bf J_{mc}} dV \times {\bf R}}{4\pi R^3} 
       = -\frac{\mu_0 I_{mc} d{\bf l} \times {\bf R}}{4\pi R^3} \ .
\label{Eele}
\end{equation}
Now we are ready to solve the electromagnetic field of our 1D moving magnetic
moment lattice because it is equivalent to two lines of positive/negative 
moving magnetic ``charges'': they are easily obtained by integrating 
Eqs.(\ref{B0}) and (\ref{Eele}) respectively. The same final results of
Eqs.(2,\ref{E1}) are obtained. The present derivation allows 
us to conclude that the magnetic field ${\bf B}$ is induced by magnetic 
moment and the electric field ${\bf E}$ is induced by the magnetic 
moment current. Fig.1f shows electric field lines and magnetic field 
lines at the y-z plane, here the infinitely long magnetic moment lattice 
is along the x-axis and $\hat{n}_m$ is along the +z-direction. 

If there exists another infinitely long magnetic moment lattice with 
opposite magnetic moment direction ($-\hat{n}_m$) and opposite moving 
direction ($-{\bf v}$, shown in Fig.1a), then the net magnetic moment is 
canceled exactly and only a net magnetic moment current exists. In this case, 
it is easy to confirm that the magnetic field ${\bf B}$ due to each
lattice adds up to zero identically, while the electric field ${\bf E}$ 
reinforce each other so that the total electric field of the composite system
is doubled. Hence we conclude that this finite electric field must originate 
from the magnetic moment current, and it cannot be due any other effects.

In the example above, we have clearly shown that a moving classical 1D 
magnetic moment
can induce an electric field ${\bf E}$. In the following we investigate the 
question: can moving electron magnetic moment ({\it i.e.} spin) induce an 
electric field? We also extend the above 1D model to general situation.
Before proving this is indeed the case, we emphasize the fact that 
since a magnetic moment (or a spin) is itself a vector unlike charge 
which is a scalar, the magnetic moment current density cannot be 
described only by a single vector ${\bf J}_m dV$ (or $I_m d{\bf l}$). 
In order to completely describe a magnetic 
moment current density, we have to use a set of two vectors ($\hat{n}_m$, 
${\bf J}_m dV$) or ($\hat{n}_m$, $I_m d{\bf l}$), in which ${\bf J}_m$ 
expresses the strength and direction of the flow of magnetic moment current, 
while $\hat{n}_m$ expresses the polarization of the magnetic moment itself. 
This is different from the familiar charge current.  Note that for two 
magnetic moment current such as that of Fig.1a, if only their ${\bf J}_m$ are 
the same and their $\hat{n}_m$ are different, they are two different magnetic 
moment currents and their induce electric fields are also different (see below).

In the following, we apply the equivalent magnetic ``charge'' method to deduce 
a general result beyond 1D for the quantum object of electron spin-current. 
Here, the spin or magnetic moment ${\bf m}$ of an electron at space point 
${\bf r}$ is equivalent to a positive magnetic charge $\frac{m}{\delta}$ at 
${\bf r}+\frac{\delta}{2} \hat{n}_m$ and a negative magnetic charge 
$-\frac{m}{\delta}$ at ${\bf r}-\frac{\delta}{2} \hat{n}_m$.  The 
spin-current ($\hat{n}_m$, ${\bf J}_m dV$) at the space ${\bf r}$ is 
equivalent to two magnetic ``charge'' currents: one is 
$\frac{{\bf J}_m}{\delta}dV$ at ${\bf r}+\frac{\delta}{2} \hat{n}_m$ and the
other $-\frac{{\bf J}_m}{\delta}dV$ at ${\bf r}-\frac{\delta}{2} \hat{n}_m$,
where $\delta \rightarrow 0^+$. We make the very reasonable fundamental 
assumption that any electromagnetic field induced by moving electron spins, 
if exists, must satisfy relativistic covariance. From this assumption, the 
Maxwell equations for the magnetic ``charge'' and its current are:
\begin{eqnarray}
\nabla \times {\bf E} & =& -\frac{\partial {\bf B}}{\partial t}
       - \mu_0 {\bf J}_{mc} , \label{max1} \\
\nabla \times {\bf B} & =& \mu_0 \epsilon_0\frac{\partial {\bf E}}{\partial t}
       + \mu_0 {\bf J}_{ec} , \label{max2} \\
\nabla \bullet {\bf E} & =& \rho_{ec}/\epsilon_0 ,\label{max3} \\
\nabla \bullet {\bf B} & =& \mu_o \rho_{mc}, \label{max4}
\end{eqnarray}
where $\rho_{mc}$ and $\rho_{ec}$ are the volume density of the magnetic and
electric charge, ${\bf J}_{mc}$ and ${\bf J}_{ec}$ are their current density.
In contrast, in the original Maxwell equation, the source of field 
are electric charge and its current: the field of a magnetic moment is
calculated by turning this moment into an infinitesimal charge-current loop
as we have done above. Here, we use the magnetic ``charge'' description and 
its associated current to express the spin of particles and the 
spin-current. We emphasize two points: (i) Eqs.(5-8) are superiorer 
for our problem of calculating fields of electron spin-current because 
they do not require us to turn electron spins into little charge-current 
loops. No one knows how to do the latter, in fact, because the inner 
structure of an electron is not known. Hence, while the original Maxwell 
equations do not directly describe fields of electron spin and the spin-current,
Eqs.(5-8) can describe them and this description is very reasonable if 
we only investigate fields outside of an electron, {\it e.g.} for 
$R>10^{-5}$\AA. (ii) Eqs.(\ref{max1}--\ref{max4}) do not represent an 
attempt of rewriting Maxwell equation. They {\it are} the Maxwell equation 
when we use equilent magnetic ``charge'' and demanding relativistic 
covariance.  

From the Eqs.(\ref{max1}-\ref{max4}), the electric field induced by an 
infinitesimal element of magnetic ``charge'' current ${\bf J}_{mc} dV$ is 
obtained from Eq.(4). Then the total electric field ${\bf E}$ of the 
element of magnetic moment current ($\hat{n}_m$, ${\bf J}_m dV$) can be 
calculated by adding the two contributions of the two magnetic ``charge'' 
currents: $\frac{{\bf J}_m}{\delta}dV$ at $\frac{\delta}{2} \hat{n}_m$ and 
$-\frac{{\bf J}_m}{\delta}dV$ at $-\frac{\delta}{2} \hat{n}_m$ 
($\delta \rightarrow 0^+$). We obtain
\begin{equation}
{\bf E}=
 \int  \frac{\mu_0}{4\pi} {\bf J}_m dV \times \frac{1}{R^3}
   \left[ \hat{n}_m - \frac{ 3{\bf R} \left( {\bf R} \bullet \hat{n}_m
         \right) }  {R^2} \right]\ .
\label{Etotal}
\end{equation}
This is one of the main results of this paper.  Eq.(\ref{Etotal}) clearly 
shows that the magnetic moment current ($\hat{n}_m$, ${\bf J}_m dV$) 
({\it i.e.} spin-current ($\sigma$, ${\bf J}_s dV$)=
($\hat{n}_m$, $\frac{{\bf J}_m}{g\mu_B} dV$)) indeed can produce an electric
field. This formula can be thought as the `` Biot-Savart law'' for
spin-current induced electric field. We emphasize that in the derivation 
of Eq.(\ref{Etotal}), 
the only assumption made was that the electromagnetic field of the moving 
spin satisfies relativistic covariance. As a check, applying 
Eq.(\ref{Etotal}) to the 1D lattice exactly solved above, it is 
straightforward to perform the integration and obtain Eq.(\ref{E1}). 

Some further discussion of our results are in order. An electron has its 
charge and magnetic moment (spin): charge produces electric field, 
charge-current produces magnetic field, spin produces magnetic field, and we 
have just shown that a steady state spin-current produces an electric field! 
(i) For the case of a spin-current without charge-current shown in Fig.1a, 
the total net charge is zero for our neutral system; the total 
charge-current is zero; and the total magnetic moment is also zero. The 
only non-zero quantity is the total spin-current (magnetic moment current). 
Our results predict even for this situation, an electric field is induced 
by the presence of spin-current. (ii) For the spin-polarized charge-current 
shown in Fig.1b, which have been extensively investigated 
recently\cite{ref1,ref2}, a charge-current, total magnetic moment, 
and a spin-current may all exist. In this case, the charge-current and magnetic 
moment produce magnetic field, and the spin-current produces electric field.
(iii) For a closed-loop circuit in which a steady state spin-current flows, 
one can prove that the induced electric field ${\bf E}$ has the property 
$\ointop_c {\bf E} \bullet d{\bf l} =0$, where $C$ is an arbitrary close 
contour not cutting the spin-current. This is true even when the spin-current 
threads the contour $C$: very different from the Ampere's law of magnetic 
field induced by a charge-current.

Fig.2 shows electric field lines of a spin-current element 
($\hat{n}_m$, ${\bf J}_m dV$). The spin-current element is located at 
origin, ${\bf J}_m$ points to $+{\bf x}$ direction, and $\hat{n}_m$ is in 
the x-z plane. The angle between ${\bf J}_m$ and $\hat{n}_m$ is $\theta$. 
Because the induced electric field ${\bf E}$ must be perpendicular to 
${\bf J}_m$ ({\it i.e.} to {\bf x} axis), we plot the field lines in the 
y-z plane at $x=-1$, 0, and +1. (i) For $\theta=\pi/2$, 
$\hat{n}_m \perp {\bf J}_m$. At $x=0$, the field line configuration is similar 
(although not exactly the same) to that produced by an electric dipole 
at $-{\bf y}$ direction (Fig.2a). At $x=\pm 1$, the field lines have a 
mirror symmetry between upper and lower half y-z plane (Fig.2b). 
(ii) For $\theta=0$, $\hat{n}_m \parallel {\bf J}_m$. 
At $x=0$, ${\bf E}=0$ for any y and z. At $x=\pm 1$, the field lines 
are concentric circles (Fig.2c and d). The center of the circles is
at $y=z=0$ where  ${\bf E}=0$.  (iii) For $\theta =\pi/3$, the fields are
shown in Fig.2(e,f) for $x=\pm 1$.  In fact, this ${\bf E}$ can be decomposed into 
a summation of two terms corresponding to the fields of $\theta =\pi/2$ and 
$\theta =0$. At $x=0$, the field lines are similar to that shown in Fig.2a. 
It is worth to mention that from Fig.2, it is clearly shown that
$\ointop_c {\bf E} \bullet d{\bf l} \not= 0$. Notice that this is not 
contradictory to characteristic (iii) of the last paragraph, because there 
the electric field is produced by a steady closed-loop spin-current.

So far we have demonstrated that a spin-current can indeed induce an electric 
field. On the other hand, does external electric field have any effect on a 
spin-current?  Consider a lab frame $\Sigma '$ where there is an 
electromagnetic field $({\bf E}',{\bf B}')$, and a {\it static} magnetic 
moment ${\bf m}'$. There is a potential energy $-{\bf m}' \bullet {\bf B}'$ but 
${\bf m}'$ does not couple to ${\bf E}'$. Using the language of magnetic 
``charge'' discussed above, we can in effect consider that a force 
$q_{mc} {\bf B}'$ is acting on the magnetic ``charge'' $q_{mc}$. Inside a new 
frame $\Sigma$ moving with speed $-{\bf v}$ respect to the lab frame $\Sigma '$,
the magnetic moment ${\bf m}$ (or the magnetic charge $q_{mc}$) move with speed 
$+{\bf v}$. A Lorentz transform of the the four-momentum 
$p_{\mu} =({\bf p}, \frac{iW}{c})=(p_1,p_2,p_3, \frac{iW}{c})$ from frame 
$\Sigma'$ to $\Sigma$ gives the force ${\bf F}$ on the moving magnetic 
``charge'' $q_{mc}$ by the electromagnetic field ${\bf E},{\bf B}$:
${\bf F} =\frac{d{\bf p}}{dt} =\frac{d{\bf p}}{d\tau} \frac{d\tau}{dt}
=\left(\frac{dp_1'}{d\tau}, \gamma^{-1}\frac{dp_2'}{d\tau},
  \gamma^{-1}\frac{dp_3'}{d\tau}\right)
= q_{mc} (B_1',  \gamma^{-1}B_2', \gamma^{-1}B_3')
=q_{mc} \left({\bf B}- \frac{{\bf v}}{c^2} \times {\bf E}\right)$
where $\tau$ is the proper time, 
$t$ is the time, the quantity with a prime (without prime) is in frame 
$\Sigma'$ ($\Sigma$), and the direction of $p_1$ is the same as velocity 
${\bf v}$. Hence, a moving magnetic moment ${\bf m}$ (or spin $\sigma$) 
in an external electromagnetic field ${\bf E},{\bf B}$ feels a torque:
${\bf m}\times \left( {\bf B} -\frac{{\bf v}}{c^2}\times {\bf E}\right)$, 
and the associated potential energy is:\cite{note2}
\begin{equation}
  -{\bf m} \bullet \left( {\bf B} -\frac{{\bf v}}{c^2} \times {\bf E}\right)
 = -g\mu_B {\bf \sigma} \bullet
 \left( {\bf B} -\frac{{\bf v}}{c^2} \times {\bf E}\right)\ .
\label{force1}
\end{equation}
Clearly, the term $-{\bf m}\bullet {\bf B}$ describes the action of magnetic
field on ${\bf m}$ which is well known. There is a new term 
${\bf m} \bullet \left( \frac{{\bf v}}{c^2} \times {\bf E}\right)$, and it 
obviously expresses the action of electric field ${\bf E}$ on the {\it moving}
magnetic moment. We therefore conclude that when a particle with spin 
${\sigma}$ is moving inside an electric field ${\bf E}$, the spin prefers to
orient to the direction of $-{\bf v}\times {\bf E}$.

At last, it should be mentioned some previous works have investigated effects of
moving magnetic dipole ${\bf m}$ in early days of 
special relativity.\cite{book2}
The main finding was that a moving magnetic dipole 
induces an electric dipole ${\bf P}_e =\frac{{\bf v}}{c^2}\times {\bf m}$. 
Notice the our present work is different. What we predicted is that a {\it steady
state} spin-current can induce an electric field ${\bf E}$ (see Eq.(9)). This 
gives a new and fundamental source of electric field. 
In constract, a single moving moment does not give a steady state spin-current. 
Moreover, the results are different. For example, 
when ${\bf v} \parallel {\bf m}$, the induced dipole
${\bf p}_e =0$ so that no electric field is induced and any torque 
${\bf p}_e \times {\bf E}$ from external electric field vanishes.
Our result on spin-current induced field, on the other hand, gives a non-zero 
${\bf E}$ when ${\bf v} \parallel {\bf m}$.

So far we have found that a spin current can induce an electric field 
${\bf E}$; and conversely, an external electric field puts a moment of force 
on a spin-current. The magnitudes of these effects can be estimated.  
Consider a spin current ($\hat{n}_m$, ${\bf J}_m$) flowing in an infinitely 
long wire with crossection area of 2mm$\times$2mm. Let $\hat{n}_m \perp {\bf J}_m$, 
take electron density is $10^{29} /m^3$ and a drift velocity $10^{-2}m/s$, then 
the spin-current induced electric field is equivalent to that of a potential 
difference $\sim 12\mu$V at distances $-1.1$mm and $1.1$mm on either side
of the wire. This electric potential is indeed very small, but is definitely nonzero
and should be measurable using present technologies.

{\bf Acknowledgments:} 
We thank Prof. D. Stairs for an illuminating discussion about electromagnetism.
We gratefully acknowledge financial support from Natural Science and
Engineering Research Council of Canada, le Fonds pour la Formation de
Chercheurs et l'Aide \`a la Recherche de la Province du Qu\'ebec, and
NanoQuebec (Q.S., H.G,),
and a RGC grant from the
SAR Government of Hong Kong under grant number HKU 7091/01P (J.W.).

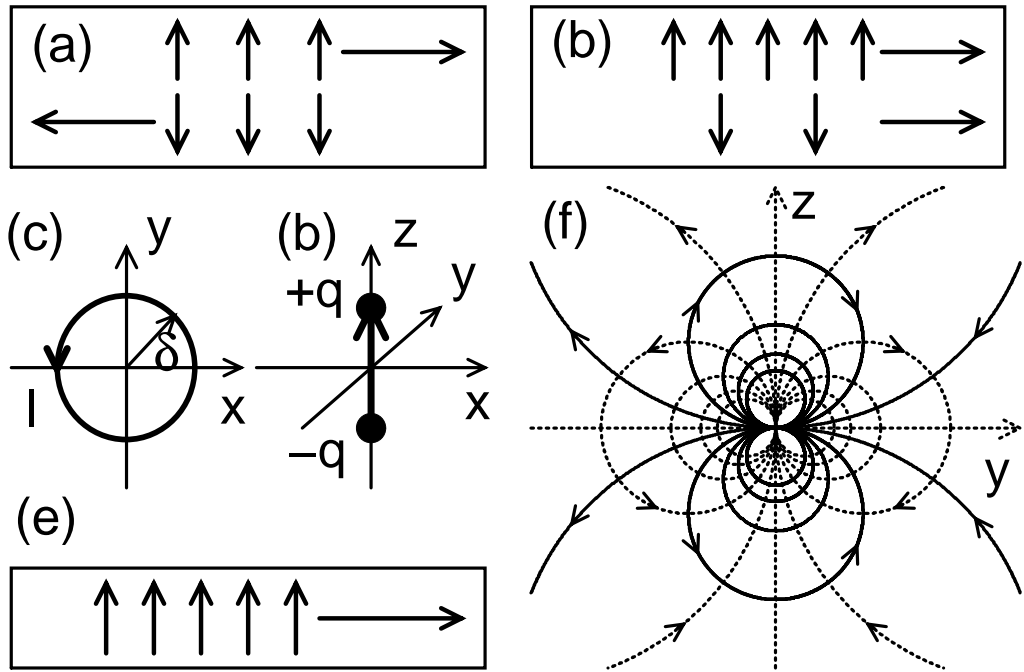
\begin{figure}
\caption{
Schematic plots for the spin current with zero charge current (a), the
spin-polarized current (b), the magnetic moment of a small current ring (c), 
the two equivalent magnetic charges of the magnetic moment (d), and the 
straight infinite long magnetic moment line (e). f shows the electric (solid) 
and magnetic (dotted) line of force for the motive magnetic moment line.
}
\label{fig1}
\end{figure}

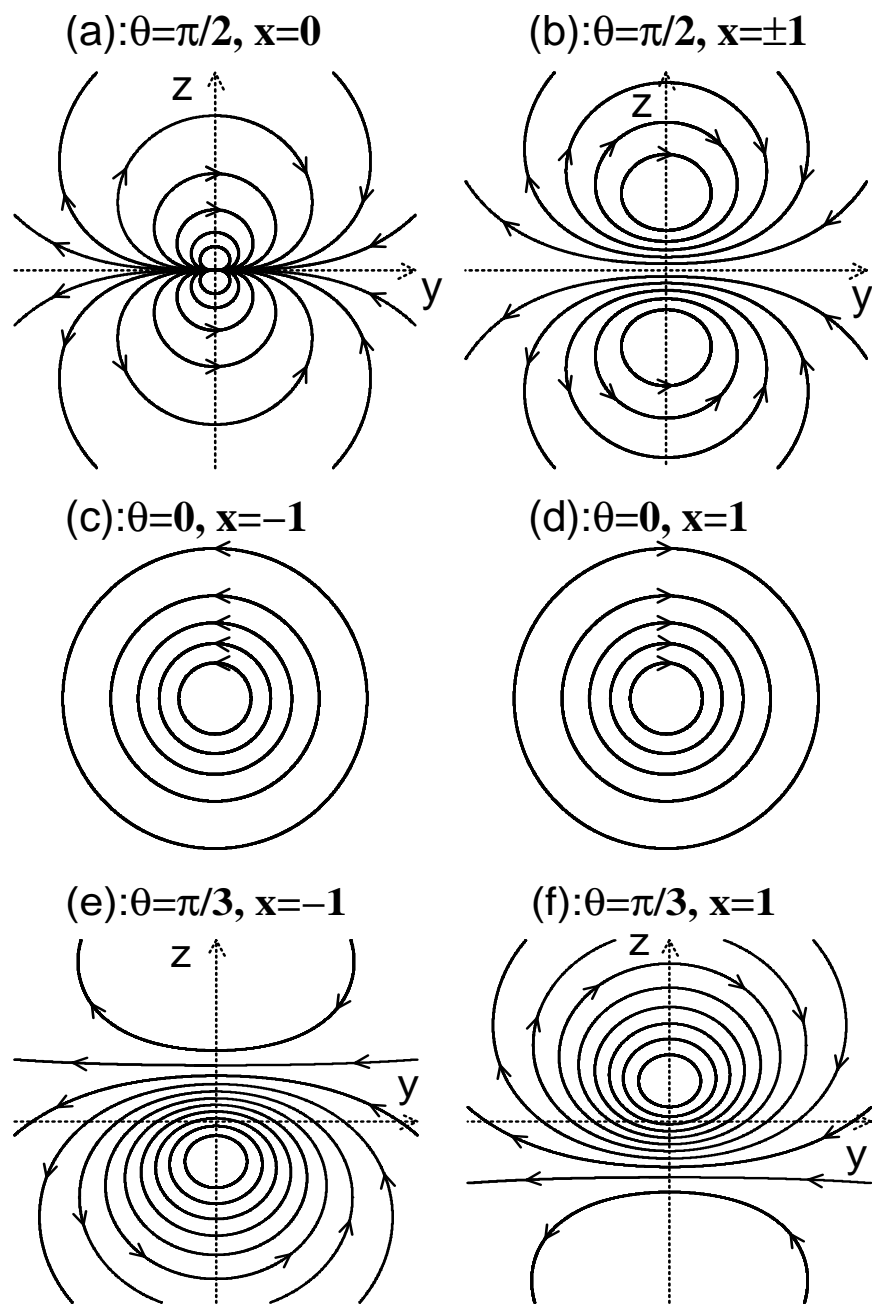
\begin{figure} 
\caption{ 
Schematic plots of electric field lines for a spin-current 
element with $\theta =\pi/2,\pi/3$ and $0$.
} 
\label{fig2}
\end{figure}

\begin{references}
\bibitem{note2}
Notice where the electromagnetic field ${\bf E}$ and ${\bf B}$ are nonuniform,
the motive magnetic moment ${\bf m}$ except have the moment of force
${\bf m} \times \left( {\bf B} -\frac{{\bf v}}{c^2} \times {\bf E}\right)$,
it too has the net force $({\bf m}\bullet \nabla) \left( {\bf B}
-\frac{{\bf v}}{c^2} \times {\bf E}\right)$.
\end{references}
\end{document}